\documentclass[reprint,superscriptaddress,amssymb,amsmath,prd,aps,showpacs,nofootinbib]{revtex4-1}%

\usepackage{graphicx}
\usepackage{bm}
\usepackage{mathtools}
\usepackage{dcolumn}
\usepackage{natbib}
\usepackage{color}
\usepackage{upgreek}
\usepackage{siunitx}
\usepackage{hyperref}
\hypersetup{colorlinks,
    citecolor = blue}

\newcolumntype{m}{D{+}{\,\pm\,}{-1}}
\DeclareSIUnit\parsec{pc}
\newcommand{\ud}{\text{d}}
\newcommand{\thr}{\text{th}}
\newcommand{\cmb}{\text{CMB}}
\newcommand{\obs}{\text{obs}}
\newcommand{\jla}{\text{JLA}}

\newcommand{\bao}{\text{BAO}}
\newcommand{\CC}{\text{CC}}
\newcommand{\C}{\mathbf{C}}
\newcommand{\D}{\mathbf{D}}

\newcommand{\AIC}{\text{AIC}}
\newcommand{\BIC}{\text{BIC}}
\newcommand{\maxi}{\text{max}}
\newcommand{\mini}{\text{min}}
\newcommand{\bmdmu}{\bm{\Delta \mu}}

\begin{document}

\title{Cosmic chronometers constraints on some fast-varying dark energy equations of state}

\author{Rafael J. F. Marcondes}
\email{rafaelmarcondes@usp.br}
\affiliation{Departamento de F\'isica Matem\'atica, Instituto de F\'isica, Universidade de S\~ao Paulo, Rua do Mat\~ao 1371, S\~ao Paulo, Brazil}

\author{Supriya Pan}
\email{span@iiserkol.ac.in}
\affiliation{Department of Mathematics, Raiganj Surendranath Mahavidyalaya, Sudarshanpur, Raiganj, West Bengal 733134, India}
\affiliation{Department of Physical Sciences, Indian Institute of Science Education and Research Kolkata, Mohanpur, West Bengal 741246, India}

\pacs{98.80.-k, 95.36.+x, 95.35.+d, 98.80.Es}

\begin{abstract}
We consider three `four-parameters' dark energy equations of state  allowing fast transition
from the matter dominated decelerating phase to the current accelerating phase. The fast-varying nature of the dark energy 
models is quantified by the transition width
$\tau > 0$, a free parameter associated with the models where lower values of $\tau$ imply faster transition. We impose the latest observational constraints on
these fast-varying dark energy equations of state, using the latest released
cosmic chronometers data along with a series of standard dark energy probes, namely, 
the local Hubble constant value at \SI{2.4}{\percent} precision measured by the
Hubble Space Telescope, the Joint Light Curve Analysis from Supernovae Type Ia,
Baryon acoustic oscillations distance measurements and finally the cosmic microwave
background radiation distance priors. Our analyses show that the precise measurements of the free parameters, when a large number of parameters are allowed in a cosmological model become very hard. Moreover, the analyses do not enable us to make any decisive comment on the fast-varying nature of the models, at least from the astronomical data available at current moment. Finally, we close the work with a discussion based on the information criteria, which do not return favorable results to the fast-varying models, at least according to the data employed.   
\end{abstract}

\maketitle


\section{Introduction}

The discovery of the current cosmic acceleration has thrown us into 
an inordinate challenging phase of modern cosmology. This period 
remains as one of the strangest episodes of the dynamical 
history of the universe yet. Several theoretical proposals have been recommended 
in the last couple of years aiming to explain this observed accelerating
phase. The simplest of such proposals is to introduce some dark energy
fluid in the context of Einstein's general theory of relativity.
The dark energy is some kind of artificial fluid that arises
from the modifications of the matter sector of the universe
when gravity is described by the general relativity
\cite{copeland2006,amendola-dark-2010, Bamba:2012cp}. Aside from the
concept of dark energy in Einstein gravity, this accelerating universe can also
be realized in modified gravity models that either appear from the simple
extension (s) of the Einstein-Hilbert action
\cite{Capozziello:2002rd, Nojiri:2003ft, Carroll:2003wy, Das:2005bn, de-felice-fr-2010,Sotiriou:2008rp, Capozziello:2011et, nojiri-unified-2011,Nojiri:2017ncd,capozziello-cauchy-2011} (also see \cite{Capozziello:2007ec, Capozziello:2005ku, Amarzguioui:2005zq, Capozziello:2006dj, Paliathanasis:2011jq, Bel:2014awa, Paliathanasis:2016tch, Nunes:2016drj, Odintsov:2017qif}) or with the
teleparallel gravitational theory \cite{cai-f-2016, Geng:2011aj, Li:2011wu, deHaro:2012zt, Nesseris:2013jea, Paliathanasis:2016vsw, Nunes:2016qyp, Bahamonde:2017bps, Awad:2017yod}. 
However, in this
work we shall consider the dark energy models. Now, within the dark energy
candidates, the most elementary and the simplest dark energy
candidate is the cosmological constant, $\Lambda$, with equation of state
$w_{\Lambda} = P_{\Lambda}/\rho_{\Lambda} = -1$, that together with cold
dark matter (CDM) offers an excellent description to the present
accelerated expansion of the universe. However, despite being so successful with
the astronomical observations, theoretical issues like the fine tuning \cite{Weinberg-1988} and the
coincidence problems \cite{Zlatev-1999} unveil its limitations.  
Thus, a variety of alternative dark energy models, from 
non-minimally coupled scalar field models \cite{Ratra-1987} to more complicated cosmological models \cite{Nojiri-2005}, have been introduced so far. For a review 
on different dark energy models and their effects on the dynamical universe, we further refer to \citet{copeland2006,amendola-dark-2010,Bamba:2012cp}.

Along the same direction of research, parametric dark
energy models came into existence where the equation of
state for the dark energy model, namely, $w_d (z)= P_d/\rho_d$, is expressed 
in terms of the redshift parametrizations. Thus, with a given functional form for $w_d (z)$,
the expansion history of the universe can be reconstructed. Finally, using the astronomical data from various potential sources, the viabilities and the limitations of such models are 
tested.
The widely known dark energy parametrizations
are linear parametrization 
\cite{cooray-gravitational-1999,astier-can-2001,weller-future-2002},
Chevallier-Polarski-Linder parametrization 
\cite{chevallier-accelerating-2001,linder-exploring-2003}, logarithmic
parametrization \cite{efstathiou-constraining-1999}, 
Jassal-Bagla-Padmanabhan (JBP) parametrization 
\cite{jassal-wmap-2004}, Barboza-Alcaniz parametrization 
\cite{Barboza-2008} and many more, see for instance 
\cite{Nesseris:2005ur,Wang:2008zh,Ma:2011nc,Feng:2012gf,Zhang:2015lda,Pantazis:2016nky,Yang:2017alx}.
The common behavior in 
the above mentioned models is that they contain only two free
parameters.
Similarly, one can extend the two-parameters 
family of models into three-parameters family of 
dark energy parametrizations \cite{Wetterich:2004pv,Ichikawa:2006qt,Liu:2008vy,Lazkoz:2010gz,linder-how-2005}
as well as to four-parameters family of dark energy
parametrizations \cite{Corasaniti-2001,Bassett-2002,linder-how-2005,felice-observational-2012}.

In the current work we focus on some four-parameters family of the dark energy
parametrizations 
\citep{Bassett-2002,Corasaniti-2001,felice-observational-2012}
which allow a fast transition from the past decelerating expansion 
to the present accelerating expansion and impose an updated observational 
constraints on the models using the latest released cosmic chronometers 
data set \cite{moresco-6-2016} along with 
some standard dark energy probes, namely, the Joint-light 
Curve analysis \cite{betoule-improved-2014} from Supernoave Type Ia (SNIa), baryon acoustic oscillations distance measurements \cite{beutler-6df-2012,ross-clustering-2015,anderson-clustering-2014,kazin-wigglez-2014,font-ribera-quasar-lyman-2014}, 
cosmic microwave background radiation \cite{planck-collaboration-planck-2016} and the local Hubble constant value \cite{riess-2.4-2016} from the Hubble Space Telescope (HST). We perform a robust statistical analysis using the Markov Chain Monte Carlo (MCMC) method to extract the information out of the cosmological models. 
The background geometry is described by the Friedmann-Lema\^itre-Robertson-Walker (FLRW) line element, as usual.

We organize this work in the following way. After briefly discussing the
field equations in FLRW universe in section \ref{sec-bckg-eq}, we introduce three well known
fast-varying dark energy parametrizations in Section \ref{sec:models}.  Section
\ref{sec-data} deals with different observational data that we used. In section
\ref{sec-results} we have described the results from the observational analysis, which we also use to compare the viability of the models in section \ref{sec:aic-bic}.
Finally, section \ref{sec-discu} concludes the main summary of the work.

\section{Background equations}
\label{sec-bckg-eq}
On the largest scales our universe is perfectly homogeneous and isotropic and
this is characterized by the FLRW line element
\begin{align}
    \ud s^2= -\ud t^2+ a^2(t) \left[\frac{\ud r^2}{1-\kappa r^2} + r^2 \left(\ud \theta^2 + \sin^2 \theta \, \ud \phi^2 \right)  \right], 
\end{align}
where $a(t)$ is the expansion scale factor of the universe; $\kappa$ is the spatial curvature which for $0$, $+1$, $-1$ represents  respectively a flat, closed and an open universe. In such a background, the Friedmann equations are 
\begin{align}
H^2 + \frac{\kappa}{a^2} &= \frac{8 \pi G}{3} \left(\rho_r+ \rho_b + \rho_c + \rho_d  \right),\\
2 \dot{H}+  3 H^2 + \frac{\kappa}{a^2} &= -8 \pi G \left(P_r + P_b + P_c +P_d  \right),
\end{align}
where an overhead dot represents the differentiation with respect to the cosmic time, $t$; $H \equiv \dot{a}/a$ is the Hubble expansion rate; $\rho_r$, $\rho_b$, $\rho_c$ and
$\rho_{d}$ are, respectively, the energy densities of radiation, baryons, cold
dark matter and dark energy (DE), while $P_r$, $P_b$, $P_c$ and $P_d$ are the pressures of the corresponding sectors, and also we assume that the equation of state of dark energy is barotropic, i.e.
$w_{d} = P_{d}/\rho_{d}$.
Since from the latest observations \cite{planck-collaboration-planck-2016} the
spatial curvature of the universe is almost zero, thus, in agreement with the
observational data as well as for simplicity we assume $\kappa =0$ in this
work.
Now, since all the fluids are non-interacting, they obey the usual conservation
law $\dot{\rho}_i+ 3 H (P_i+ \rho_i)= 0$. 
In particular, the evolution of DE fluid is governed by 
\begin{align}
    \label{cons}
    \rho_{d}= \rho_{d0}\,\exp\left(-
        \int_{1}^{a}\frac{3\left(1+w_d\right)}{\tilde{a}}\,\ud\tilde{a}\right).
\end{align}
Thus, for a spatially flat FLRW universe one can write down the first Friedmann
equation as 
\begin{align}
    \label{Friedmann}
    \frac{H^2}{H_0^2} =  \frac{\Omega_{r0}}{a^4} + \frac{\Omega_{m0}}{a^3} +
 \Omega_{d0}\,e^{-\int_{1}^{a}3\left(1+w_d\right)\tilde{a}^{-1}\,\ud\tilde{a}},
\end{align}
where $\Omega_{r0}+ \Omega_{m0}+ \Omega_{d0} = 1$ and $\Omega_{m0} =
\Omega_{b0} + \Omega_{c0}$.

\section{Models}
\label{sec:models}

Dark energy models which allow a fast transition from the past matter dominated
decelerating phase to current observed acceleration are termed as fast varying
dark energy equations of state. In order to quantify their fast varying nature,
one needs to increase the number of free parameters into the dark energy
equation of state. In this section we shall present some dark energy
parametrizations that were studied in
\cite{Bassett-2002,Corasaniti-2001,felice-observational-2012}. 

\subsection{Model 1}

We introduce the first fast-varying dark energy parametrization
\cite{linder-how-2005} which takes the following form 
\begin{align}
    \label{model1}
    w_d (a) = w_f + \frac{w_p - w_f}{1+ \left(a / a_t\right)^{1/\tau}},
\end{align}
where $a_t$ is the scale factor at the transition era, that means this is the
value of the expansion scale factor at which the universe entered into the
current accelerating phase from the decelerating matter dominated era; $\tau~(>
0)$ is the transition width, physically which means the time elapsed by the model to enter the accelerating regime from the past decelerating phase; $w_f$, $w_p$ are the free parameters of this model where $w_f = \lim_{a\rightarrow \infty} w_d(a)$, and $w_p = \lim_{a\rightarrow 0} w_d (a)$. Now, solving the conservation equation (\ref{cons}) for this model and then using it into the Friedmann  equation (\ref{Friedmann}) one has 
\begin{align}
    \frac{H^2}{H_0^2}= \frac{\Omega_{r0}}{a^4} + \frac{\Omega_{m0}}{a^3} + \frac{\Omega_{d0}}{a^{3\left(1+w_p\right)}} f_1 (a),
\end{align}
where the function $f_1 (a)$ is the following 
\begin{equation}
    f_1 (a) = \left(\frac{a^{1/\tau}+ a_t^{1/\tau}}{1+ a_t^{1/\tau}} \right)^{3
        \tau \left(w_p- w_f \right)}.
\end{equation}
One can see that the model (\ref{model1}) does not allow any extremum of $w_d
(a)$ for any value of the scale factor. This was reported by
\citet{felice-observational-2012} and the authors resolved such extremum
problems with some new models. In the following we introduce such models to
extract the observational constraints using the Markov Chain Monte Carlo
simulations.

\subsection{Model 2}

Let us introduce a second model which was first proposed in
ref.~\cite{felice-observational-2012}
\begin{align}
    \label{model2}
    w_d (a) = w_p + \left(w_0 - w_p\right) a \frac{1-\left(a/a_t \right)^{1/\tau}}{1-\left(1/a_t\right)^{1/\tau}}.
\end{align}
where $w_0$ is the current value of $w_d (a)$ and $a_t$, $\tau (>0)$, $w_p$ have the same meanings as described for Model 1 in (\ref{model1}). Similarly, one can solve the conservation
equation (\ref{cons}) for this dark energy equation of state and finally the
Friedmann equation (\ref{Friedmann}) takes the form 
\begin{align}
    \frac{H^2}{H_0^2}= \frac{\Omega_{r0}}{a^4} + \frac{\Omega_{m0}}{a^3} + \frac{\Omega_{d0}}{a^{3\left(1+w_p\right)}} \, e^{f_2 (a)},
\end{align}
where the function $f_2 (a)$ is given by 
\begin{align}
    f_2 &(a) = 3\left(w_0- w_p\right) \times {} \nonumber \\
    {} &\times \frac{1+ \bigl(1-a_t^{-1/\tau}\bigr)\tau + a \bigl[\bigl\lbrace\left(a/a_t\right)^{1/\tau}-1 \bigr\rbrace\tau -1 \bigr]}{\left(1+\tau\right) \bigl(1-a_t^{-1/\tau}\bigr)}
\end{align}
One can verify that the model (\ref{model2}) admits an extremum at $a_{\ast} =
\left[\tau/(\tau+1)\right]^{\tau} a_t$, where the dark energy equation of
state is \cite{felice-observational-2012}
\begin{eqnarray}
w_{\ast} = w_{d} (a_{\ast}) = w_p + \frac{(w_0-w_p) \, \tau^{\tau} a_t^{1+1/\tau}}{(1+\tau)^{1+\tau} \bigl( a_t^{1/\tau}- 1\bigr)}~.
\end{eqnarray}
For more analysis in this direction we refer to
\citet{felice-observational-2012}.

\subsection{Model 3}

Finally, the last model in this work follows \cite{felice-observational-2012}
\begin{align}
    \label{model3}
    w_d (a) = w_p + \left(w_0 - w_p\right) a^{1/\tau}
    \frac{1-\left(a/a_t \right)^{1/\tau}}{1-\left(1/a_t\right)^{1/\tau}}.
\end{align}
where all the model parameters, namely, $a_t$, $\tau~(> 0)$, $w_0$ and $w_p$ have the same interpretation as described for Model 2 in (\ref{model2}). Using (\ref{model3}) into
(\ref{cons}), the Friedmann equation (\ref{Friedmann}) for this model can be solved as  
\begin{align}
    \frac{H^2}{H_0^2}=  \frac{\Omega_{r0}}{a^4} + \frac{\Omega_{m0}}{a^3} + \frac{\Omega_{d0}}{a^{3\left(1+w_p\right)}} \, e^{f_3 (a)},
\end{align}
where the function $f_3 (a)$ has the following expression
\begin{align}
    f_3 (a) = 3 &\left(w_0- w_p\right) \tau  \times {} \nonumber \\
    {} & \times \frac{2-a_t^{-1/\tau}+ a_t^{1/\tau}
        \bigl[\left(a/a_t\right)^{1/\tau}-2\bigr]}{2 \bigl(1-a_t^{-1/\tau}\bigr)}
\end{align}
Similar to the previous model (\ref{model2}), this model also has the extremum
\cite{felice-observational-2012} at $a_{\ast} = a_t/2^\tau$, 
where the dark energy equation of state takes the value 
\begin{eqnarray}
    w_{\ast} = w_d( a_{\ast}) = w_p + \frac{1}{4} \left[ \frac{\left(w_0
                -w_p \right) a^{1/\tau}}{a^{1/\tau}- 1} \right].
\end{eqnarray}
A discussion on the nature of the extremum is given by
\citet{felice-observational-2012}.

\section{Observational data}
\label{sec-data}
Our analyses combine data from different probes.
We detail below how we calculate the likelihoods for the cosmic chronometer (CC)
dataset with the local measurement of $H_0$, Type Ia supernovae (JLA binned
data), baryon acoustic oscillation (BAO) data and CMB distance priors.
The combined total likelihood $\mathcal{L}$ will be given by $\log \mathcal{L} = 
    \log \mathcal{L}_{\CC} +
    \log \mathcal{L}_{H_0}+
    \log \mathcal{L}_{\jla} +
    \log \mathcal{L}_{\bao} + 
    \log \mathcal{L}_{\cmb}$,
which is summed with the log-prior probability to give the log-posterior probability.
We employ a MCMC code to carry out a Bayesian parameter inference for the models of fast-varying equation of state.

\subsection{Cosmic chronometer dataset}
\label{cc-data}

The cosmic chronometer approach is a method to determine the Hubble parameter values at different redshifts with the use of most massive and passively evolving galaxies.
These galaxies are known as cosmic chronometers.
The method calculates $\ud z/\ud t$ and hence the Hubble parameter using the relation $H(z)= - (1+z)^{-1} \ud z/\ud t$. 
Since the measurement of $\ud z$ is obtained through spectroscopic method with high accuracy, a precise measurement of the Hubble parameter lies on the precise measurement of the differential age evolution $\ud t$ of such galaxies, and hence these measurements are considered to be model independent.
A detailed description about the cosmic chronometer method can be found in ref.~\cite{moresco-6-2016}.
Here we use the \num{30} measurements of the Hubble parameter in the redshift
interval $0 < z< 2$  \cite{moresco-6-2016}, which are listed in table~\ref{tab:CCdata}. 
\begin{table}[tb]
    \caption{\label{tab:CCdata}Cosmic chronometer data from ref.~\cite{moresco-6-2016}.}
    \begin{ruledtabular}
        \begin{tabular}{D..{-1} m  D..{-1} m}
            \mbox{$z$}     &      \mbox{$H(z)$}  & \mbox{$z$}     &      \mbox{$H(z)$} \\ \hline
            0.07  &   69.0+19.6  & 0.4783 & 80.9+9 \\
            0.09  &   69+12 & 0.48  &   97+62 \\
            0.12  &   68.6+26.2 & 0.593  &  104+13 \\
            0.17  &   83+8 & 0.68  &   92+8 \\
            0.179  &  75+4 & 0.781  &  105+12 \\
            0.199  &  75+5 & 0.875  &  125+17 \\
            0.20  &   72.9+29.6 & 0.88  &   90+40 \\
            0.27  &   77+14 & 0.9  &    117+23 \\ 
            0.28  &   88.8+36.6 & 1.037  &  154+20 \\
            0.352  &  83+14 & 1.3  &    168+17 \\
            0.3802 & 83+13.5 & 1.363  &  160+33.6 \\
            0.4  &   95+17 & 1.43  &   177+18 \\
            0.4004  & 77+10.2 & 1.53  &   140+14 \\
            0.4247  & 87.1+11.2 & 1.75  &   202+40 \\
            0.44497 & 92.8+12.9 & 1.965  &  186.5+50.4
        \end{tabular}
    \end{ruledtabular}
\end{table}

The likelihood for the cosmic chronometer data is calculated as
\begin{align}
    \label{likelihood-cc}
    2 \log \mathcal{L}_{\CC} = - 30 \log(2 \uppi) - \sum_{i=1}^{31} \left[ 2 \log \sigma_i + \chi_{\CC,i}^2 \right],
\end{align}
with $\chi^2_{\CC,i} = \bigl[ H(z_i)^{(\obs)} - H(z_i)^{(\thr)} \bigr]^2 /
\sigma_i^2$, where the $\sigma_i$ are the uncertainties in the $H(z)$
measurements for each data point $i = 1, \dots, 30$.

\subsection{Local Hubble constant value}

We also include the local value of the Hubble parameter
directly measured from the luminosity distances by Riess 
et al.~\cite{riess-2.4-2016}. The local Hubble constant value yields
$H_0^{(\obs)} \pm \sigma_{H_0} = \SI{73.24 \pm 1.74}{\km\per\s\per\mega\parsec}$ with \SI{2.4}{\percent}
precision (we denote this value as R16). The likelihood for this datum is
calculated as
\begin{align}
    2 \log \mathcal{L}_{H_0} = - \log(2 \uppi) - 2 \log \sigma_{H_0} - \chi_{H_0}^2,
\end{align}
with $\chi^2_{H_0} = \bigl[ H_0^{(\obs)} - 100 \,h \, \si{\km\per\s\per\mega\parsec} \bigr]^2 /
\sigma_{H_0}^2$, where the $\sigma_{H_0}$ is the uncertainty in the measurement
of the local Hubble constant.

\subsection{Type Ia Supernovae}
\label{snia-data}

The acceleration of the expansion of the universe was discovered by using luminosity distances of Type Ia supernovae (SNe Ia) as standard candles \cite{riess-observational-1998,perlmutter-measurements-1999}. 
We use the estimates of binned distance modulus $\mu_b$ obtained from the joint
analysis of the SDSS-II and SNLS supernova catalogues---the JLA sample,
consisting of 31 points (30 bins) \cite{betoule-improved-2014}, given in
table~\ref{tab:JLAdata}. 
\begin{table}[tb]
    \caption{\label{tab:JLAdata}Binned distance modulus fitted to the JLA
        sample. From ref.~\cite{betoule-improved-2014}.}
    \begin{ruledtabular}
        \begin{tabular}{D..{-1} D..{-1}  D..{-1} D..{-1}}
            \mbox{$z_b$}     &      \mbox{$\mu_b$}  & \mbox{$z_b$}     &      \mbox{$\mu_b$} \\ \hline
            0.010  &  32.954 &    0.134  &  39.068   \\
            0.012  &  33.879 &    0.158  &  39.341  \\
            0.014  &  33.842 &    0.186  &  39.792  \\
            0.016  &  34.119 &    0.218  &  40.157  \\
            0.019  &  34.593 &    0.257  &  40.565  \\
            0.023  &  34.939 &    0.302  &  40.905  \\
            0.026  &  35.252 &    0.355  &  41.421  \\
            0.031  &  35.749 &    0.418  &  41.791  \\
            0.037  &  36.070 &    0.491  &  42.231  \\
            0.043  &  36.435 &    0.578  &  42.617  \\
            0.051  &  36.651 &    0.679  &  43.053  \\
            0.060  &  37.158 &    0.799  &  43.504  \\
            0.070  &  37.430 &    0.940  &  43.973  \\
            0.082  &  37.957 &    1.105  &  44.514  \\
            0.097  &  38.253 &    1.300  &  44.822  \\
            0.114  &  38.613 &           &
        \end{tabular}
    \end{ruledtabular}
\end{table}
These data serve as a good approximation to the full JLA likelihood. 
In a flat universe, the distance modulus $\mu_{\thr}$ of an object is
\begin{align}
    \mu_{\thr}(z) = 5 \log_{10} \frac{d_L (z)}{\SI{10}{\parsec}},
\end{align}
where $z$ is the redshift of the supernova and the luminosity distance $d_L$ is given by
\begin{align}
    d_L(z) = \left(1+z\right) \frac{c}{H_0} \int_0^{z} \frac{\ud \tilde
        z}{E(\tilde z)},
\end{align}
where $c$ is the speed of light and $E(z) \equiv H(z) / H_0$.

We compare the observed distance modulus $\mu_b$ with the theoretical value
$\mu_{\thr}$ plus a shift parameter $M$ that we marginalize over.
The approximated likelihood $\mathcal{L}_{\jla}$ is given by
\begin{align}
    2 \log \mathcal{L}_{\jla} = - 31 \log(2 \uppi) - \log(\det \C) - \chi_{\jla}^2
\end{align}
with the chi-square function
\begin{align}
    \chi_{\jla}^2 = \bmdmu^{\mathsf{T}} \C^{-1} \bmdmu,
\end{align}
where $\bmdmu \equiv {\bm \mu}_{b} - \left({\bm \mu}_{\thr} + \mathbf{M} \right)$, is
the data comparison vector for all supernovae and $\mathbf{M}$ the column vector with
all values equal to $M$.
$\C$ is the covariance matrix for the binned data given by
\citet{betoule-improved-2014}.

\subsection{Baryon Acoustic oscillation}
\label{bao-data}
Characteristic scales left in the matter distribution can be detected in galaxy surveys and in the Ly$\alpha$ forest emission lines of distant quasars. In our analysis we include the data measurements of baryon acoustic oscillations (BAO) from different surveys:
the Six Degree Field Galaxy Survey (6dF) \cite{beutler-6df-2012},
the Main Galaxy Sample of Data Release 7 of Sloan Digital Sky Survey (SDSS-MGS) \cite{ross-clustering-2015},
the LOWZ and CMASS galaxy samples of the Baryon Oscillation Spectroscopic Survey
(BOSS-LOWZ and BOSS-CMASS) \cite{anderson-clustering-2014},
the WiggleZ Dark Energy Survey \cite{kazin-wigglez-2014}
and the distribution of the Lyman $\alpha$ forest in BOSS (BOSS-Ly)
\cite{font-ribera-quasar-lyman-2014}.
The data are listed in table~\ref{tab:BAOdata} and give a measurement of the ratio $r_{\bao} \equiv r_s(z_d)/d_V(z)$ between the sound horizon 
\begin{align}
    \label{eq:soundhorizon}
    r_s(z) = \frac{c}{H_0} \int_z^{\infty} \frac{\ud \tilde z}{E(\tilde z) \sqrt{3 \left[1 + \frac{3}{4} \frac{\Omega_{b0}}{\Omega_{r0}} \frac{1}{1 + \tilde z} \right]}}
\end{align}
when baryons were released from the Compton drag of photons \cite{eisenstein-baryonic-1998},
\begin{align}
    z_d = \frac{1291 \left(\Omega_{m0} h^2 \right)^{0.251}}{1 + 0.659 \left(\Omega_{m0} h^2 \right)^{0.828}} \left[ 1 + b_1 \left( \Omega_{b0} h^2 \right)^{b_2} \right],
\end{align}
and the effective BAO distance \cite{eisenstein-detection-2005}
\begin{align}
    d_V(z) \equiv \left[ \frac{\left(1 + z\right)^2 \left\lbrace d_a(z) \right\rbrace^2 c z}{H(z)} \right]^{1/3},
\end{align}
where 
$b_1 = 0.313 \left( \Omega_{m0} h^2 \right)^{-0.419} \left[1 + 0.607 \left( \Omega_{m0} h^2 \right)^{0.674} \right]$, 
$b_2 = 0.238 \left( \Omega_{m0} h^2 \right)^{0.223}$ and
\begin{align}
    \label{eq:angdiadistance}
    d_a(z) = \frac{1}{1 + z} \frac{c}{H_0}  \int_0^z \frac{\ud \tilde z}{E(\tilde z)},
\end{align}
is the angular diameter distance.
\begin{table}[tb]
    \caption{\label{tab:BAOdata}BAO data measurements included in our analysis.}
    \begin{ruledtabular}
        \begin{tabular}{l D..{-1}  m  c}
            Survey      &       \mbox{$z$} & \mbox{$r_{\bao}$}   &     Reference \\ \hline
            6dF         &       0.106   &      0.336+0.015   & \cite{beutler-6df-2012} \\
            SDSS-MGS    &       0.15  &     0.224+0.008 &   \cite{ross-clustering-2015} \\
            BOSS-LOWZ   &       0.32  &     0.1181+0.0023 &   \cite{anderson-clustering-2014} \\
            WiggleZ     &       0.44  &     0.0888+0.0043   &   \cite{kazin-wigglez-2014} \\
            BOSS-CMASS  &       0.57  &     0.0726+0.0007    &   \cite{anderson-clustering-2014} \\
            WiggleZ     &       0.6   &     0.0686+0.0031    &   \cite{kazin-wigglez-2014}  \\
            WiggleZ     &       0.73   &    0.0605+0.0021    &   \cite{kazin-wigglez-2014} \\
            BOSS-Ly$\alpha$ &   2.36  &     0.033+0.001   &   \cite{font-ribera-quasar-lyman-2014}
        \end{tabular}
    \end{ruledtabular}
\end{table}
The last data point is actually a combination of two measurements which are the BAO scale along the line of sight $c/H(z) r_s(z_d) = \num{9.0 \pm 0.3}$ and across the line of sight $d_a(z)/r_s(z_d) = \num{10.8 \pm 0.4}$, both at $z = 2.36$ \cite{font-ribera-quasar-lyman-2014}.
Noting that
\begin{align}
    r_{\bao} = \frac{r_s(z_d)}{d_V(z)} = \left[ \frac{H(z) r_s(z_d)/c}{z \left(1 + z\right)^2 \left\lbrace d_a(z)/r_s(z_d) \right\rbrace^2}
\right]^{1/3},
\end{align}
the two measurements combined yield the ratio $r_{\bao}(z=2.36) = \num{0.033 \pm 0.001}$.

The likelihood is calculated as
\begin{align}
    2 \log \mathcal{L}_{\bao} = - 8 \log(2 \uppi) - \sum_{i=1}^8 \left[2 \log
        \sigma_i + \chi_{\bao,i}^2 \right],
\end{align}
where $\chi_{\bao,i}^2 = \bigl[r_{\bao,i}^{(\obs)} - r_{\bao,i}^{(\thr)} \bigr]^2/\sigma_i^2$ and $\sigma_i$'s are the uncertainties in the measurements for each data point $i = 1, \dots, 8$.

\subsection{Cosmic Microwave Background data}
We use distance priors obtained from Planck TT, TE, EE + lowP data assuming a $w$CDM cosmology \cite{huang-distance-2015} given in terms of three parameters: an acoustic scale $l_A$, a shift parameter $R$ and the amount of baryons $\Omega_{b0} h^2$. 
The acoustic scale is
\begin{align}
    l_A \equiv \frac{1}{\theta_A} = \left( 1 + z^* \right) \uppi \, \frac{d_a(z^*)}{r_s(z^*)},
\end{align}
where $\theta_A$ is the observation angle subtending the transverse comoving scale $\lambda_p = r_s(z^*)/\uppi$ of the first acoustic peak, $r_s$ is the sound horizon given by eq.~\eqref{eq:soundhorizon}, $d_a$ the angular diameter distance from eq.~\eqref{eq:angdiadistance}, $z^*$ is the redshift to the photo-decoupling surface, given by the fitting formula \cite{hu-smallscale-1996} 
\begin{align}
    z^* = \num{1048} \left[ 1 + 0.00124 \left( \Omega_{b0} h^2 \right)^{-0.738} \right] \left[ 1 + g_1 \left( \Omega_{m0} h^2 \right)^{g_2} \right],
\end{align}
with $g_1 = \frac{0.0783 \left( \Omega_{b0} h^2 \right)^{-0.238}}{1 + 39.5 \left( \Omega_{b0} h^2 \right)^{0.763}}$ and $g_2 = \frac{0.560}{1 + 21.1 \left( \Omega_{b0} h^2 \right)^{1.81}}$.
The shift parameter is
\begin{align}
    R(z^*) \equiv \left(1 + z^*\right) \frac{ d_a(z^*) }{c/H_0}\sqrt{\Omega_{m0}} = \sqrt{ \Omega_{m0} } \int_0^{z^*} \frac{\ud z}{E(z)}.
\end{align}

The observed data are $R = 1.7488 \pm 0.0049$, $l_A = 301.498 \pm 0.091$ and $\Omega_{b0} h^2 = 0.02228 \pm 0.00016$.
With these errors and the correlation matrix
\begin{align}
    {\bm \rho}  = \begin{bmatrix}
            1.0 &   0.49    &   -0.68 \\
            0.49    &   1.0 &   -0.38 \\
            -0.68   &   -0.38   &   1.0 \\
        \end{bmatrix},
\end{align}
a covariance matrix $\D$ with $\D_{ij} = {\bm \rho}_{ij} \sigma_i \sigma_j$,
$i,j= R, l_A, \Omega_{b0} h^2$ is also given,
so the CMB chi-square is given by $\chi^2_{\cmb} = \mathbf{X}^{\mathsf{T}} \D^{-1} \mathbf{X}$, with $\mathbf{X} = \left( R^{\thr} - 1.7488; l_A^{\thr} - 301.498; \Omega_{b0} h^2 - 0.02228 \right)$ and the log-likelihood is
\begin{align}
    2 \log \mathcal{L}_{\cmb} = - 3\log(2 \uppi) - \log\left( \det \D\right) - \chi^2_{\cmb}.
\end{align}

\section{Results}
\label{sec-results}
In the following we present the results of our MCMC analyses for the three models using the combined analysis CC + JLA + BAO + CMB + R16. The convergence of the chains was analyzed according to
the multivariate extension by \citet{brooks-general-1998} of the method by
\citet{gelman-inference-1992}.
This consists of monitoring a between-chain covariance $\hat{V}$ and a
within-chain covariance $W$, determining convergence when a distance measurement
$\hat{R}^p$ between $\hat{V}$ and $W$ indicates that they are satisfactorily close.
A proper measurement of $\hat{R}^p$ is given by the square root of the maximum
eigenvalue of the positive definite matrix $W^{-1} \hat{V}$ and since it should
be close to 1 we determine convergence when $\bigl| \hat{R}^p - 1 \bigr|$ is
smaller than some precision $\epsilon$.

\begin{table*}[t]
    \caption{\label{tab:fv1results}Constraints on the parameters of the
        fast-varying equation of state parametrization of Model 1
        from the combined analysis CC + JLA + BAO + CMB + R16.}
    \begin{ruledtabular}
        \renewcommand{\arraystretch}{1.2}
        \begin{tabular}{c c D..{-1} m m}
            Parameter	 & 	Prior	 & 	\multicolumn{1}{c}{Best-fit}	 & \multicolumn{1}{c}{$1\sigma$ C.L.}	 & 	\multicolumn{1}{c}{$2\sigma$ C.L.}  \\ \hline
            $h$	&	$\left[0.50 ,0.90\right]$	&	0.70801	&	0.70968+0.01249	 &	0.70968+0.02498	\\
            $\Omega_{c0}h^2$	&	$\left[0.09 ,0.17\right]$	&	0.13701	&	0.13666+0.00591	 &	0.13666+0.01182	\\
            $100\,\Omega_{b0}h^2$	&	$\left[1.80 ,2.40\right]$	&	2.22069	&	2.22394+0.01590 	&   2.22394+0.03181	\\
            $10^{5}\,\Omega_{r0}h^2$	&	$\left[3.00 ,8.50\right]$	&	5.76349	&	5.75924+0.24898	&	5.75924+0.49796	\\
            $w_p$	&	$\left[-2.00 ,1.00\right]$	&	-0.17641	&	\multicolumn{1}{c}{$-0.74414^{+0.51650}_{-0.26131}$}	&	\multicolumn{1}{c}{$-0.74414^{+0.89927}_{-0.74550}$}	\\
            $w_f$	&	$\left[-4.00 ,0.00\right]$	&	-1.08577	&	\multicolumn{1}{c}{$-1.04854^{+0.09090}_{-0.35669}$}	&	\multicolumn{1}{c}{$-1.04854^{+0.20805}_{-0.86894}$}	\\
            $a_t$	&	$\left[0.00 ,1.00\right]$	&	0.30133	&	\multicolumn{1}{c}{(unconstrained)}	&	\multicolumn{1}{c}{(unconstrained)}	\\
            $\tau$	&	$\left[0.00 ,1.00\right]$	&	0.14413	&	\multicolumn{1}{c}{$\tau \geqslant 0.44$}	&	\multicolumn{1}{c}{$\tau \geqslant 0.09$}	\\
            $\Delta M$ (nuisance)	&	$\left[-0.30 ,0.30\right]$	&	0.01920	&	\multicolumn{1}{c}{$0.02350^{+0.03637}_{-0.04164}$}	&	\multicolumn{1}{c}{$0.02350^{+0.07766}_{-0.08073}$}	\\
            \hline
            $\Omega_{c0}$	&		&	0.27332	&	0.27131+0.00898	&	0.27131+0.01797	\\
            $100\,\Omega_{b0}$	&		&	4.43007	&	4.41985+0.15976	&   4.41985+0.31952	\\
            $10^{5}\,\Omega_{r0}$	&		&	11.49761	&	11.43426+0.37779	&	11.43426+0.75558	\\
            $\Omega_{d0}$	&		&	0.68227	&	0.68437+0.00937     &	0.68437+0.01874	\\            
        \end{tabular}
    \end{ruledtabular}
\end{table*}
\begin{figure*}[bt]
    \centering
    \includegraphics[width=\textwidth]{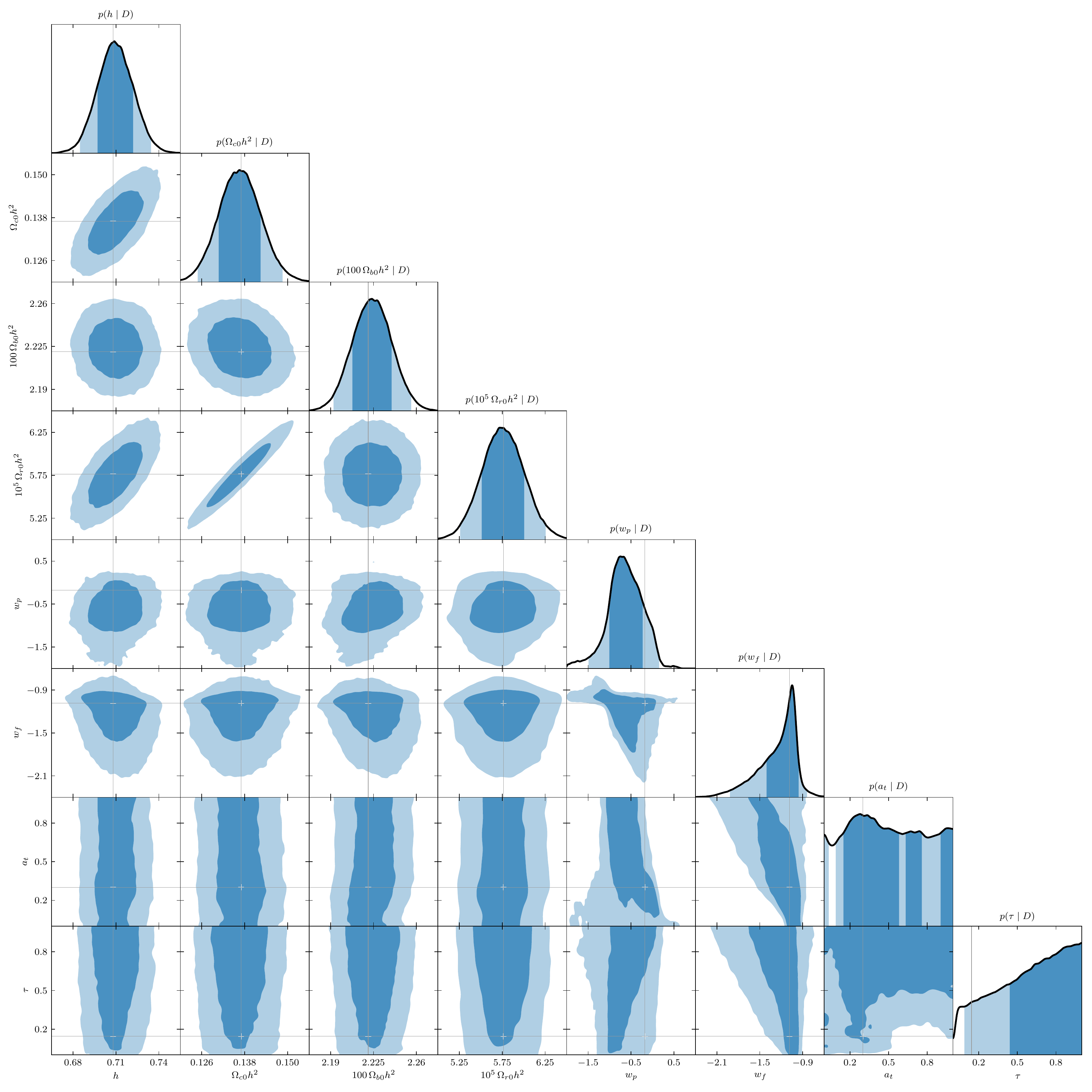}
    \caption{\label{fig:fv1_joint}Results of the fast-varying EoS
        parametrization of Model 1. Light and dark blue shaded areas mark the
        \SI{95}{\percent} and \SI{68}{\percent} ($2\sigma$ and $1\sigma$)
        confidence level regions, respectively. The crosses and the vertical
        dashed lines indicate the values of the parameters at the best-fit
        point.}
\end{figure*}

\subsection{Model 1}
Starting with Model 1, we now present its results
extracted using the combined analysis mentioned above.
The results, given in table~\ref{tab:fv1results}, were obtained from a standard 
MCMC simulation with twelve chains.
Convergence within $|\hat{R}^p - 1| < 0.03$ between the chains
was achieved. Better convergence with reasonable chain sizes seems to be
prevented, in this case, by the difficulty in sampling the posterior
distribution along the $a_t$-axis. 
This parameter is totally unconstrained by the data. The analysis also shows that the model has a tendency to approch the cosmological constant limit,  while the constraints on $\tau$ are not so small, at least from the present observational data that we employ.  Constraints on the derived parameters $\Omega_{c0}$, $\Omega_{b0}$,
$\Omega_{r0}$ and $\Omega_{d0}$ are also given in table~\ref{tab:fv1results}. Figure~\ref{fig:fv1_joint} displays the marginalized posterior probability
distributions and the contour levels of the two-parameter joint posterior probabilities for this model.

\subsection{Model 2}
\begin{figure*}[bt]
    \centering
    \includegraphics[width=\textwidth]{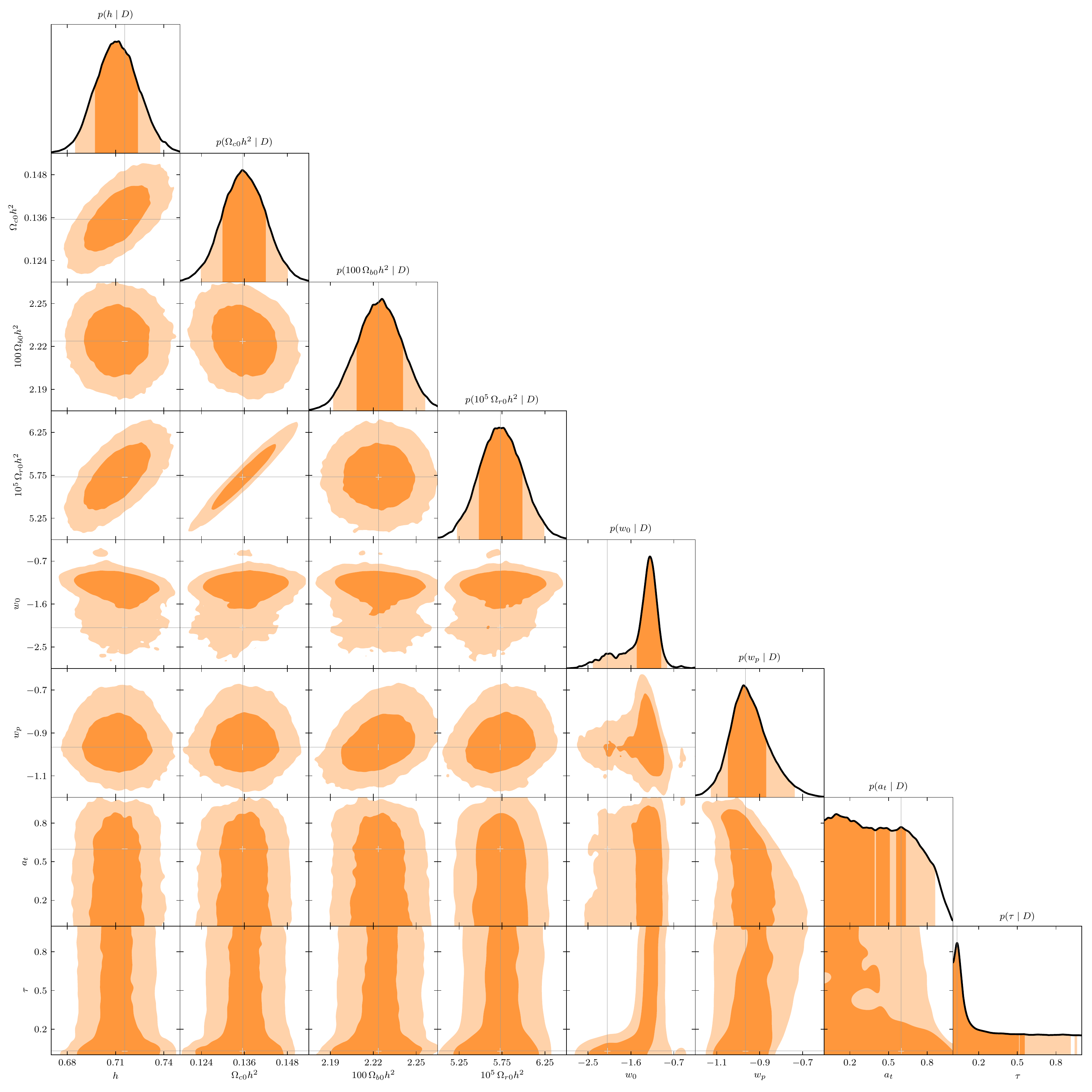}
    \caption{\label{fig:fv2_joint}Results of the fast-varying EoS
        parametrization of Model 3. Light and dark orange shared areas mark the
        \SI{95.4}{\percent} and \SI{68.3}{\percent} confidence level regions,
        respectively. The thin grey lines indicate the values of the parameters
        at the best-fit point.}
\end{figure*}

The prior ranges and the results of the analysis of the second model are given
in table~\ref{tab:fv2results}. The marginalized posterior probability
distributions and the contour levels of the two-parameter joint posterior
probabilities are presented in figure~\ref{fig:fv2_joint}.
These results were obtained from a standard MCMC simulation with five chains,
achieving convergence $|\hat{R}^p - 1| < 0.03$ after \num{830000} steps, of which the first half is discarded.
Constraints on the derived parameters $\Omega_{c0}$, $\Omega_{b0}$,
$\Omega_{m0}$, $\Omega_{r0}$ and $\Omega_{d0}$ as well as the scale
\begin{align}
    a_* = \left( \frac{\tau}{\tau + 1} \right)^{\tau} a_t
\end{align}
at which $w_d$ has the extremum
\begin{align}
    w_* = w_p + \left(w_0 - w_p\right)  \frac{\tau^{\tau} \left(\tau+1\right)^{-\tau-1} a_t}{1 - a_t^{-1/\tau}}
\end{align}
are also given in table~\ref{tab:fv2results}. From the analysis it is clear that $a_t$ is slightly better constrained than in Model 1, with its maximum limit in the $1\sigma$ C.L.~ constrained to be $a_t \lesssim 0.60$, which does not imply the transition at recent past, as argued by the current observational data. However, the constraint on $\tau$ is relatively small and within $1\sigma$ C.L., $\tau =0$ is allowed. This result is in favor of a fast-varying dark energy model. Additionally, we also find that the current value of the dark energy equation of state has phantom nature. 
\begin{table*}[t]
    \caption{\label{tab:fv2results}Constraints on the parameters of the
        fast-varying equation-of-state parametrization of Model 2 extracted
        from the combined analysis CC + JLA + BAO + CMB + R16.}
    \begin{ruledtabular}
        \renewcommand{\arraystretch}{1.2}
        \begin{tabular}{c c D..{-1} m m}
            Parameter    &    Prior    &    \multicolumn{1}{c}{Best-fit}    &    \multicolumn{1}{c}{$1\sigma$ C.L.}    &    \multicolumn{1}{c}{$2\sigma$ C.L.}    \\    \hline
            $h$	&	$\left[0.50 ,0.90\right]$	&	0.71565	&	0.71107+0.01326 &	0.71107+0.02652	\\
            $\Omega_{c0}h^2$	&	$\left[0.06 ,0.20\right]$	&	0.13551	&	0.13593+0.00606	&	0.13593+0.01212	\\
            $100\,\Omega_{b0}h^2$	&	$\left[1.00 ,4.00\right]$	&	2.22372	&	2.22422+0.01617	&	2.22422+0.03234	\\
            $10^{5}\,\Omega_{r0}h^2$	&	$\left[3.00 ,8.50\right]$	&	5.72965	&	5.73609+0.25439	&	5.73609+0.50878	\\
            $w_0$	&	$\left[-3.00 ,0.00\right]$	&	-2.09256	&	\multicolumn{1}{c}{$-1.20103^{+0.23689}_{-0.27722}$}	&	\multicolumn{1}{c}{$-1.20103^{+0.36106}_{-1.19470}$}	\\
            $w_p$	&	$\left[-3.00 ,1.00\right]$	&	-0.96653	&	\multicolumn{1}{c}{$-0.97523^{+0.10578}_{-0.07271}$}	&	\multicolumn{1}{c}{$-0.97523^{+0.23821}_{-0.15298}$}	\\
            $a_t$	&	$\left[0.00 ,1.00\right]$	&	0.59804	&	\multicolumn{1}{c}{$a_t \lesssim 0.60$}	&	\multicolumn{1}{c}{$a_t \lesssim 0.86$}	\\
            $\tau$	&	$\left[0.00 ,1.00\right]$	&	0.03086	&	\multicolumn{1}{c}{$0.03170^{+0.52135}_{-0.03170}$}	&	\multicolumn{1}{c}{$0.03170^{+0.90239}_{-0.03170}$}	\\
            $\Delta M$ (nuisance)	&	$\left[-0.30 ,0.30\right]$	&	0.00828	&	0.01261+0.04115	&	0.01261+0.08230	\\
            \hline
            $\Omega_{c0}$	&		&	0.26459	&	0.26884+0.00962	&	0.26884+0.01924	\\
            $100\,\Omega_{b0}$	&		&	4.34184	&	4.40365+0.16796	&	4.40365+0.33592	\\
            $10^{5}\,\Omega_{r0}$	&		&	11.18722	&	11.34452+0.39746	&	11.34452+0.79491	\\
            $\Omega_{d0}$	&		&	0.69188	&	0.68701+0.01011 	&	0.68701+0.02021	\\
            $a_*$	&		&	0.53666	&	\multicolumn{1}{c}{$0.08090^{+0.31160}_{-0.07350}$}	&	\multicolumn{1}{c}{$0.08090^{+0.61329}_{-0.09051}$}	\\
            $w_*$	&		&	-0.96653	&	\multicolumn{1}{c}{$-0.96493^{+0.11867}_{-0.06685}$}	&	\multicolumn{1}{c}{$-0.96493^{+0.24481}_{-0.14584}$}	\\
        \end{tabular}
    \end{ruledtabular}
\end{table*}
\begin{table*}[t]
    \caption{\label{tab:fv3results}Constraints on the parameters of the
        fast-varying equation-of-state parametrization of Model 3
        extracted from the combined analysis CC + JLA + BAO + CMB + R16.}
    \begin{ruledtabular}
        \renewcommand{\arraystretch}{1.2}
        \begin{tabular}{c c D..{-1} m m}
                Parameter    &    Prior    &    \multicolumn{1}{c}{Best-fit}    &    \multicolumn{1}{c}{$1\sigma$ C.L.}    &    \multicolumn{1}{c}{$2\sigma$ C.L.}    \\    \hline
                $h$	&	$\left[0.50 ,0.90\right]$	&	0.70447	&	\multicolumn{1}{c}{$0.70439^{+0.10737}_{-0.03683}$}	&	\multicolumn{1}{c}{$0.70439^{+0.19561}_{-0.09129}$}	\\
                $\Omega_{c0}h^2$	&	$\left[0.06 ,0.20\right]$	&	0.12916	&	0.13680+0.00578	&	0.13680+0.01157	\\
                $100\,\Omega_{b0}h^2$	&	$\left[1.00 ,4.00\right]$	&	2.22523	&	2.22110+0.01594	&	2.22110+0.03188	\\
                $10^{5}\,\Omega_{r0}h^2$	&	$\left[3.00 ,8.50\right]$	&	5.45078	&	5.75233+0.24356	&	5.75233+0.48713	\\
                $w_0$	&	$\left[-3.00 ,0.00\right]$	&	-2.65452	&	\multicolumn{1}{c}{$-1.03975^{+0.30893}_{-1.18201}$}	&	\multicolumn{1}{c}{$-1.03975^{+1.03278}_{-1.60879}$}	\\
                $w_p$	&	$\left[-3.00 ,1.00\right]$	&	-0.98157	&	\multicolumn{1}{c}{$-1.02398^{+0.12228}_{-0.08091}$}	&	\multicolumn{1}{c}{$-1.02398^{+0.41183}_{-0.13625}$}	\\
                $a_t$	&	$\left[0.00 ,1.00\right]$	&	0.98232	& \multicolumn{1}{c}{$0.45317^{+0.25384}_{-0.28170}$}	&	\multicolumn{1}{c}{$a_t \leqslant 0.85639$}	\\
                $\tau$	&	$\left[0.00 ,1.00\right]$	&	0.03207	&	\multicolumn{1}{c}{(unconstrained)}	&	\multicolumn{1}{c}{(unconstrained)}	\\
                $\Delta M$ (nuisance)	&	$\left[-0.30 ,0.30\right]$	&	-0.01412	&	0.02226+0.03894	&	0.02226+0.07788	\\
                \hline
                $\Omega_{c0}$	&		&	0.26027	&	\multicolumn{1}{c}{$0.27491^{+0.02034}_{-0.07496}$}	&	\multicolumn{1}{c}{$0.27491^{+0.09004}_{-0.11508}$}	\\
                $100\,\Omega_{b0}$	&		&	4.48390	&	\multicolumn{1}{c}{$4.43503^{+0.37603}_{-1.15847}$}	&	\multicolumn{1}{c}{$4.43503^{+1.49609}_{-1.74618}$}	\\
                $10^{5}\,\Omega_{r0}$	&		&	10.98346	&	\multicolumn{1}{c}{$11.47833^{+0.97438}_{-3.01530}$}	&	\multicolumn{1}{c}{$11.47833^{+3.93669}_{-4.75215}$}	\\
                $\Omega_{d0}$	&		&	0.69478	&	\multicolumn{1}{c}{$0.68041^{+0.08697}_{-0.02342}$}	&	\multicolumn{1}{c}{$0.68041^{+0.13254}_{-0.10750}$}	\\
                $a_*$	&		&	0.96073	&	\multicolumn{1}{c}{$0.23662^{+0.19877}_{-0.21775}$}	&	\multicolumn{1}{c}{$0.23662^{+0.49434}_{-0.24376}$}	\\
                $w_*$	&		&	-0.98157	&	\multicolumn{1}{c}{$-1.02398^{+0.12228}_{-0.08091}$}	&	\multicolumn{1}{c}{$-1.02398^{+0.41183}_{-0.13625}$}	\\
        \end{tabular}
    \end{ruledtabular}
\end{table*}
\begin{figure*}[bt]
    \centering
    \includegraphics[width=\textwidth]{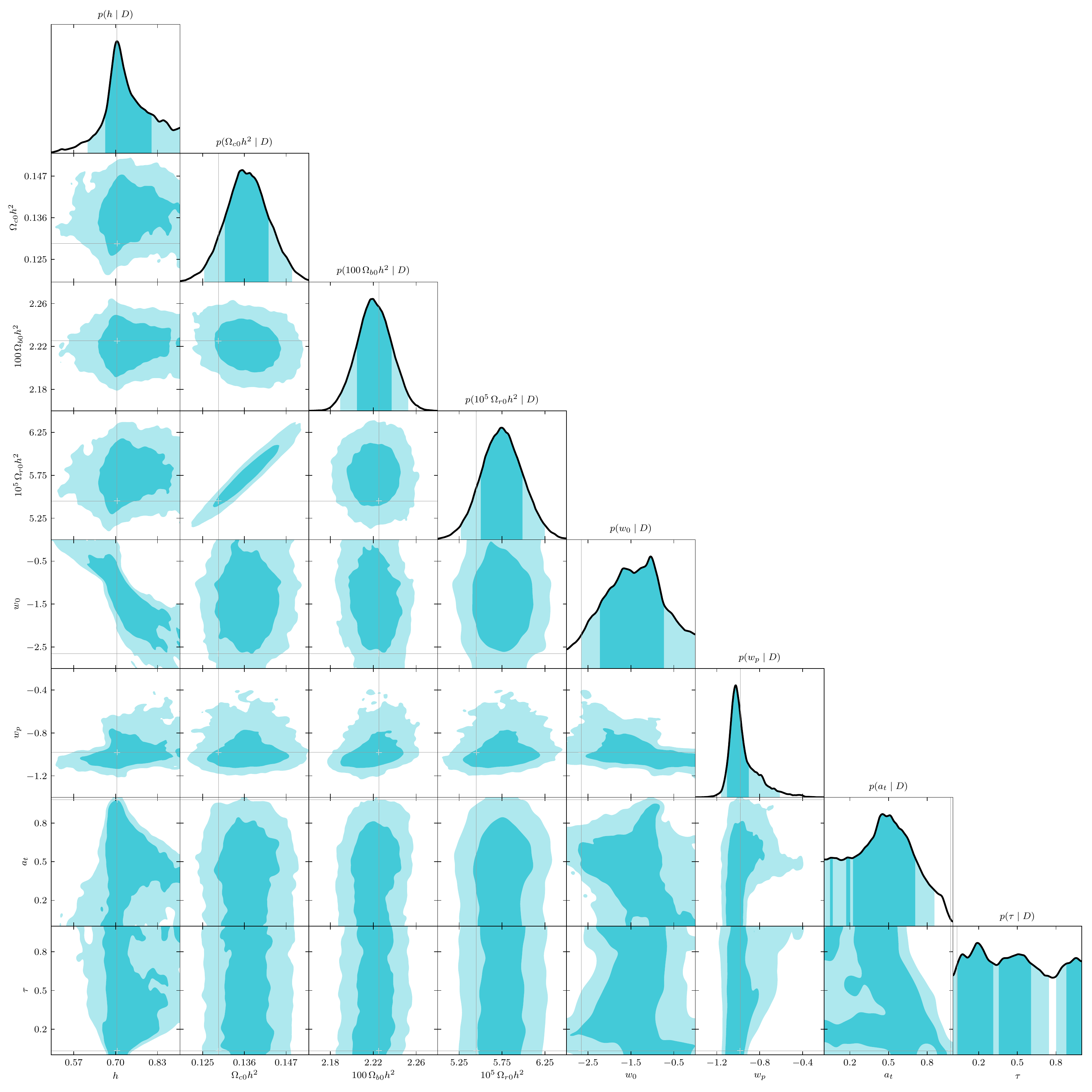}
    \caption{\label{fig:fv3_joint}Results of the fast-varying EoS
        parametrization of Model 3. Light and dark teal shared areas mark the
        \SI{95.4}{\percent} and \SI{68.3}{\percent} confidence level regions,
        respectively. The thin grey lines indicate the values of the parameters
        at the best-fit point.}
\end{figure*}

\subsection{Model 3}
The results shown in table~\ref{tab:fv3results} were obtained from a MCMC
simulation with five chains achieving convergence $| \hat{R}^p - 1 | \sim
\num{e-2}$, despite the poor sampling of the posterior distribution along the
$\tau$-axis, whose values are unconstrained. Notice also
the weak constraint on $a_t$ and even $h$ presents a very heavy tail in the
right side of the prior range. The current value of the dark energy equation of state is of phantom nature as seen from the Table \ref{tab:fv3results}.
The marginalized distributions of the relevant parameters are plotted in
figure~\ref{fig:fv3_joint}.
The results for the marginalized distributions of the derived parameters
$\Omega_{c0}$, $\Omega_{b0}$, $\Omega_{m0}$, $\Omega_{r0}$ and $\Omega_{d0}$ are
given in table~\ref{tab:fv3results}, as well as the scale
\begin{align}
    a_* = \frac{a_t}{2^{\tau}}
\end{align}
at which $w_d$ has the extremum
\begin{align}
    w_* = w_d(a_*) = w_p + \frac{1}{4} \frac{\left( w_0 - w_p \right) a_t^{1/\tau}}{1 - a_t^{-1/\tau}}.
\end{align}

\section{Information criteria}
\label{sec:aic-bic}

Finally, we close our observational analysis with the Akaike Information Criterion ($\AIC$)
\cite{akaike-new-1974} and the Bayesian (or Schwarz) Information Criterion ($\BIC$) \cite{schwarz-estimating-1978}.
The information criteria assess the viability of a cosmological model against a given reference model, under the observational considerations.
They are defined as follows:
\begin{align}
    \AIC  = -2 \ln \mathcal{L}_{\maxi} + 2 d = \chi^2_{\mini} + 2 d + C \label{aic}
\end{align}
and
\begin{align}
    \BIC  = -2 \ln \mathcal{L}_{\maxi} +  d \ln N = \chi^2_{\mini} + d \ln N + C, \label{bic}
\end{align}
where $\mathcal{L}_{\maxi}$ is the maximum value of the likelihood function, 
$d$ is the number of model parameters, $N$ is the total number of data
points used in our observational analysis and $C$ is a constant.
So, for each particular model, one can calculate AIC and BIC.
The viabilities of such models are measured by taking as reference the $\Lambda$CDM model, which is quite sound with the current observations and therefore the obvious choice.
For any given model $\mathcal{M}$, the difference $\Delta X= X_{\mathcal{M}}- X_{\Lambda\text{CDM}}$ (where
$X=\AIC$ or $\BIC$) quantifies the viability of the model. Here, 
$\Delta X > 5$ and $\Delta
X > 10$ stand respectively for strong and decisive evidences against the 
cosmological model $\mathcal{M}$ while $\Delta X <4$ is in favour with respect to the base model \cite{liddle-information-2007}. In table \ref{tab-aic-bic} we display the values of $\Delta \AIC$ and $\Delta \BIC$ for the three fast-varying dark energy models with respect to the $\Lambda$CDM cosmological model. From both information criteria summarized in Table \ref{tab-aic-bic}, it is evident that Model 1 and Model 3 have strong (from $ \Delta \AIC$) and decisive (from $\Delta \BIC$) evidences against their viabilities, while Model 2 survives from the $\AIC$ analysis but has decisive evidences against its viability from the $\BIC$ analysis.

\begin{table}[t]
\caption{\label{tab-aic-bic} Information criteria values for the three fast varying dark energy models in compared to the flat $\Lambda$CDM model. }
\begin{ruledtabular}
\renewcommand{\arraystretch}{1.2}
\begin{tabular}{ccccccccc}
Models &~~~~~ $\AIC$ &~~~~~ $\Delta \AIC$ &~~~~~ $\BIC$ &~~~~~ $\Delta \BIC$ \\
\hline
$\Lambda$CDM &~~~~~  $62.67$ &~~~~~  $0$ &~~~~~ $71.83$ &~~~~~ $0$\\

Model 1 &~~~~~  $68.16$ &~~~~~  $5.49$ &~~~~~ $86.49$ &~~~~~ $14.65$\\

Model 2 &~~~~~  $67.03$ &~~~~~  $4.35$ &~~~~~ $85.35$ &~~~~~ $13.52$\\

Model 3 &~~~~~  $68.01$ &~~~~~  $5.34$ &~~~~~ $86.33$ &~~~~~ $14.50$
\end{tabular}
\end{ruledtabular}
\end{table}

\section{Summary and conclusions}
\label{sec-discu}

Dark energy parametrizations allowing a fast transition from the past matter
dominated decelerated expansion to the current cosmic acceleration
are the main theme of this work. The models with fast-varying nature are qualitatively different from the usual two-parameters family of dark energy parametrizations \cite{cooray-gravitational-1999,astier-can-2001,weller-future-2002,chevallier-accelerating-2001,linder-exploring-2003,efstathiou-constraining-1999,jassal-wmap-2004,Barboza-2008,Nesseris:2005ur,Wang:2008zh,Ma:2011nc,Feng:2012gf,Zhang:2015lda,Pantazis:2016nky,Yang:2017alx}, and naturally extend the parameters space in terms of the new parameters quantifying such fast varying nature, for instance. 

In this work we consider three
fast-varying dark energy models proposed earlier in the literature respectively
in \cite{Bassett-2002} and \cite{felice-observational-2012} occupying the forms of Model 1 (\ref{model1}), Model 2 (\ref{model2}) and Model 3 (\ref{model3}) with an aim to impose an 
updated observational constraints on them using the latest cosmic chronometers data
set together with a series of standard dark energy probes, namely, the JLA from Supernovae Type Ia, BAO, CMB and local Hubble constant $H_0$ from HST measured with 2.4\% precision. The number of free parameters in all three models is four and the dynamics is considered in the spatially flat Friedmann-Lema\^itre-Robertson-Walker universe. 
Following the joint observational data CC + JLA + BAO + CMB +
R16, we list the observational constraints on three fast varying dark
energy equations of state (\ref{model1}), (\ref{model2}) and (\ref{model3}) respectively in tables \ref{tab:fv1results}, \ref{tab:fv2results} and \ref{tab:fv3results}.  While the figures \ref{fig:fv1_joint}, \ref{fig:fv2_joint} and \ref{fig:fv3_joint} respectively display the marginalized posterior probability
distributions and the contour levels of the two-parameter joint posterior probabilities. 

From the analyses it is quite evident that these fast-varying models cannot be well constrained, at least according to the present astronomical data employed in this work. In particular, the parameter $a_t$ is totally unconstrained in Model 1 while in Model 3, $\tau$ is unconstrained. But on the contrary, Model 2 is relatively better constrained with the present astronomical data. However, from the observational constraints on the transition width, $\tau$, quantifying the fast-varying nature, achieved only for Model 1 and Model 2, it is hard to come up with an inference about how fast-varying they really are. In fact, for Model 1, it is clear that $\tau$ assumes large values. Although for Model 2, $\tau$ is comparatively lower, but no decisive statement can be made out of these results. 
Additionally, according to the information criteria, both Model 1 and Model 3 show decisive evidences against their viabilities. However, although $\AIC$ slightly favors Model 2 but the $\BIC$ does not so.  

The current work puts a question mark on the number of free parameters allowed in a dark energy model. We recall a similar work \cite{linder-how-2005} where the authors argued that a dark energy model with more than two free parameters is quite hard to constrain. 
However, we must remark that a conclusive statement toward this direction is perhaps very hard only with the current observational data at the background level. 
A definitive conclusion may depend on a more profound and complete analysis taking into consideration the treatment of perturbations, their stability regimes, and next-generation observational data that will shed more light on important aspects of the dark energy from the large scale structure of the universe.

\section*{Acknowledgments}
SP was supported by the SERB-NPDF grant (File No. PDF/2015/000640), Government of India. Partial support from the Department  of  Atomic  Energy  (DAE), Government  of India, through NBHM Post-Doctoral grant  (File No. 2/40(60)/2015/R\&D-II/15420) is also gratefully remembered by SP.

\bibliographystyle{apsrev4-1}
\bibliography{refs}

\end{document}